
\magnification=\magstep1
\voffset=0.0truemm
\hoffset=0.0truemm
\hsize=6truein
\vsize=8.5truein
\raggedbottom
\baselineskip=24.0truept
\parindent=2.5em
\font\it=cmti12 at 12truept
\font\tenrm=cmr10 at 10truept
\font\tenit=cmti10 at 10truept

\centerline{\bf TWO-DIMENSIONAL BLACK HOLES AND}
\centerline{\bf PLANAR GENERAL RELATIVITY}
\bigskip
\centerline{\bf Jos\'e P. S. Lemos}
\centerline{\tenit Departamento de F\'{\i}sica, Instituto Superior T\'ecnico,}
\vskip -10.5truept
\centerline{\tenit  Av. Rovisco Pais 1, 1096 Lisboa, Portugal}
\vskip -9.0truept
\centerline{\tenit \&}
\vskip -9.0truept
\centerline{\tenit Departamento de Astrof\'{\i}sica, Observat\' orio
Nacional-CNPq,}
\vskip -10.5truept
\centerline{\tenit Rua General Jos\'e Cristino 77, 20921 Rio de Janeiro,
Brazil.
   }
\medskip
\centerline{\tenrm ABSTRACT}
\vskip -30.0truept

$$\vbox{\hsize 5truein\tenrm \noindent
The Einstein-Hilbert action with a cosmological term is used to derive
a new action in 1+1 spacetime dimensions.  It is shown that the
two-dimensional theory is equivalent to planar symmetry in General
Relativity. The two-dimensional theory admits black holes and free
dilatons, and has a structure similar to two-dimensional string
theories.  Since by construction these solutions also solve Einstein's
equations, such a theory can bring two-dimensional results into the
four-dimensional real world. In particular the two-dimensional black
hole is also a black hole in General Relativity.

 }$$

\noindent  General Relativity is thought to be the correct theory from
the largest conceivable scales  up to the Planck length, $10^{-33}{\rm
cm}$.  A place to test these tinyest radii can be found  in the late
stages of black hole evaporation. This is a non-trivial issue and one
is still probing regions where a semi-classical approximation is valid.
The two-dimensional (2D) black holes found in the context of string
theories [1,2] are being used to compute and analyze the back-reaction
of the radiation on the geometry [3,4,5]. Such
2D theories are, in principle, toy models and it is important to construct
a link with the four-dimensional (4D)  world. With this
purpose, we consider the following action,
$$S={1\over{2\pi}}\int{ d^2x \sqrt{-g} e^{-2\phi}\left( R + 2\left(\nabla\phi
\right)^2 +
4\lambda^2\right)},\eqno(1)$$
where $g$ is the determinant of the 2D metric, $R$ is the scalar
curvature, $\phi$ is a scalar field and $\lambda$ is a constant.
Equation (1) is very similar to the simplest 2D action derived by
imposing the vanish of the $\beta$ functions in string theory, and
given by,
$$S={1\over{2\pi}}\int{ d^2x \sqrt{-g} e^{-2\phi}\left( R + 4\left(\nabla\phi
\right)^2 +
4\lambda^2\right)}.\eqno(2)$$
As we will show, action (1) comes from 4D Einstein gravity and also
admits black hole solutions. One can generalize both these
actions into,
$$S={1\over{2\pi}}\int{ d^2x \sqrt{-g} e^{-2\phi}\left( R - 4\omega
\left(\nabla\phi\right)^2 +
4\lambda^2\right)},\eqno(3)$$
where $\omega$ is a parameter. Equations (1) and (2) have
$\omega=-{1\over2}$ and $\omega=-1$, respectively.  For $\omega=0$ one has
the Jackiw-Teitelboim theory [6,7], where the scalar curvature is a constant
and for $\omega\rightarrow\infty$ one has another constant curvature theory
which is the 2D closest analogue to General Relativity [8].
Equation (3) is a 2D  Brans-Dicke theory [9].

How can we obtain the action (1) from 4D General Relativity? The idea
is to reduce from $3+1$ to $1+1$ spacetime dimensions, loosing the less
possible information.  This is the case if, (i) the 3-space manifold
splits into a direct product of a 2D manifold $\Sigma_2$ with the real
line, $M_3=\Sigma_2 {\rm x} R$ and, (ii) the physics and geometry
(metric) on $\Sigma_2$ are invariant under the action of
the 3-parameter group, $G_3$, of motions, i.e., the action is an
isometry. Conditions (i) and (ii) imply planar symmetry. Thus the
infinitesimal generators of the group $G_3$ are two orthogonal
translations and one rotation about the axis. By a known theorem [10] if
a group $G_3$ of motions has spatial orbits of dimension 2, these
orbits admit orthogonal spacetime 2-surfaces. We will label the spatial
planar 2-surfaces by ($y,z$).

Now, Einstein-Hilbert action is,
$$S={1\over {16\pi}}\int{d^4x \sqrt{-g^{\left(4\right)}} \left( R^
{\left(4\right)} - 2\Lambda\right)}, \eqno(4)$$
where the superscript $^{\left(4\right)}$ denotes 4D quantities and
$\Lambda$ is the 4D cosmological constant. Also, the most general
plane-symmetric metric can be written as,
$$ds^2 = g_{ab} dx^a dx^b + e^{-2\phi}\left( dy^2 + dz^2\right), \eqno(5)$$
where $a,b=0,1$, and $g_{ab}$ and $\phi$ are functions on the spacetime
2-surfaces.  The scalar function $\phi$ is called for obvious reasons
the dilaton. From standard dimensional reduction techniques [11] on (4)
and (5), we obtain equation (1). Thus the 2D graviton-dilaton theory
given by equation (1) can be related in a very direct manner with
General Relativity. To obtain (1) we have integrated over a `spurious'
2D planar compact manifold, say a 2D planar torus, on which we have
imposed the normalization $\int{dy dz}=8$, and set
$\Lambda=-2\lambda^2$.  Adding a pure 2D cosmological term to the action
(1), ${1\over{2\pi}}\int{d^2x \sqrt{-g} l}$, $l$ a constant, is
equivalent to spherical symmetry in General Relativity [12].

Variation of (1) with respect to $g^{ab}$ and $\phi$ yields the gravitational
and dilaton field equations, respectively,
$${1\over2}G_{ab} + D_aD_b\phi - \left( D_a\phi\right)\left( D_b\phi\right) -
g_{ab}D_cD^c\phi +$$
$$\quad+ {3\over2}g_{ab}\left( D_c\phi\right)
\left(D^c\phi\right) - g_{ab}\lambda^2=T_{ab}e^{2\phi},\eqno(6)$$
$$2D_cD^c\phi - 2\left( D_c\phi\right)\left(D^c\phi\right) + R +4\lambda^2=0,
\eqno(7)$$
where $D$ represents the covariant derivative. We have add a matter term $S_m$
to
(1) such that ${ {\delta S_m}\over \delta g^{ab} } \equiv -{1\over\pi}\sqrt{-g}
T_{ab}$. In 2D the Einstein tensor is $G_{ab}\equiv0$.

To find solutions of this theory we have to exhibit explicitly the 2D
metric $g_{ab}$.  By performing a coordinate transformation we can put
$g_{ab}$ into diagonal form, $ds^2=-e^{2\nu}dt^2 + e^{2\mu}dx^2$, where
$\nu$ and $\mu$, (as well as $\phi$), are functions of the spacetime
coordinates ($t,x$). We still have the freedom to choose a gauge, any
gauge will do. We choose the unitary gauge, $\mu=0$. Then the metric is,
$$ds^2 = - e^{2\nu} dt^2 + dx^2.\eqno(8)$$
If we now look for static (exists a Killing vector
${\partial\over{\partial t}}$), vacuum ($T_{ab}=0$) spacetimes we
obtain from (6), (7) and (8)  the following three equations, (only two
of them are independent),
$$\phi_{,xx}-{3\over2}\phi_{,x}^2+\lambda^2=0, \eqno(9)$$
$${1\over 2}\phi_{,x}^2-\nu_{,x}\phi_{,x}-\lambda^2=0, \eqno(10)$$
$$-\phi_{,xx} + \nu_{,xx} +\nu_{,x}^2+\phi_{,x}^2-\nu_{,x}\phi_{,x}-2
\lambda^2=0.\eqno(11)$$

\noindent The linear, free, dilaton solution is of course of the form,
$\phi\propto x$. Equation (9) then gives,
$$\phi = -\sqrt{{2\over3}}\lambda x + {\rm constant},\eqno(12)$$
while (10) puts the metric in the form,
$$ds^2=-e^{{2\sqrt{2\over3}}\lambda x} dt^2 + dx^2.\eqno(13)$$

\noindent Given there is a free dilaton solution, string theories hint
that we should look for a black hole. The general solution of (9) is of
the form, $\phi=-{2\over3} \ln\left( A
\cosh\left(\sqrt{{3\over2}}\lambda x\right)
 +\right.$\break$\left. B \sinh\left(\sqrt{{3\over2}}\lambda x\right) \right).$
We now set $B=0$, (this can always be done if $A>\mid B \mid$).Then,
$$\phi=-{2\over3}\ln\left( \cosh\sqrt{{3\over2}}\lambda x\right) +\phi_0.
\eqno(14)$$.

Equation (10) yields the metric,
$$ds^2=-\tanh^2\left(\sqrt{{3\over2}}\lambda x\right) \cosh^{4\over3}
\left(\sqrt{{3\over2}}\lambda x \right)dt^2 + dx^2.\eqno(15)$$

The metric (15) has singularities at $x=0$, $x\rightarrow +\infty$ and
$x\rightarrow -\infty$. However, they are merely coordinate
singularities since the scalar curvature, $R=-{4\over3}\lambda^2 {
{\sinh^2\sqrt{{3\over2}}\lambda x}\over {\cosh^2\sqrt{{3\over2}}\lambda
x} }$, has a regular behavior at these ends. In fact, equations (14)
and (15) describe the geometry external to the horizon at $x=0$.  To
bypass the coordinate singularity at $x=0$, and to show that (15) is a
black hole, we display its maximal analytical extension. First, if
we specify the intermediary coordinate
$r={b^{1\over3}}{\sqrt{3\over2}}{1\over\lambda}
\cosh^{2\over3}\left(\sqrt{{3\over2}}\lambda x\right)$, we can put (15)
in the Schwarzschild gauge, $ds^2=-\left( a^2r^2 - {b\over ar}\right)
dt^2 + {dr^2\over {a^2r^2- {b\over {ar}}} }$,where $a\equiv
\sqrt{2\over3}\lambda$. For causal structure purposes one can set $b=1$.
We can then define a second intermediary
coordinate, $r_{*}={1\over a}\left\lbrack{1\over6}\ln\left(
ar-1\right)^2 - {1\over6}\ln\left( a^2r^2+
ar+1\right)+{1\over\sqrt{3}}\arctan{ {2ar+1}\over\sqrt{3}
}\right\rbrack$. This yields the metric in the conformal gauge,
$ds^2=-\left( a^2 r^2\left( r_{*}\right) - {1\over {ar\left(
r_{*}\right)}}\right)\left(-dt^2+{dr_{*}}^2\right)$. To expose
the maximal analytical extension we can now define the Kruskal null
coordinates,
$U=-{1\over\lambda}{\sqrt{2\over3}}e^{-\sqrt{3\over2}\lambda\left(
t-r_{*}\right)}$ and
$V={1\over\lambda}{\sqrt{2\over3}}e^{\sqrt{3\over2}\lambda\left(
t+r_{*}\right)}$. In Kruskal coordinates the metric takes the form
$$ds^2 = - {{1+{\left( ar-1\right)\over\left(
a^2r^2+ar+1\right)^{1\over2}}}\over ar} {\left( a^2r^2 +
ar +1\right)^{3\over2}\over { e^{\sqrt{3}\arctan{2ar+1\over\sqrt{3}
}} - {3\lambda^2\over2}UV}}dUdV,\eqno(16)$$
where $r$ is given implicitly as a function of $U$ and $V$.
The true singularity, at $ar=0$, obeys
$UV={2\over3\lambda^2}e^{3\pi\over{2\sqrt{3}}}$. This equation yields the
usual two-branched horizontal hyperbolae, the future branch
representing the black hole singularity and the past branch a naked
singularity.  Also, at $ar\rightarrow \infty$, one has $UV=
-{2\over3\lambda^2}e^{3\pi\over{2\sqrt{3}}}$, which are two vertical
hyperbolae. In addition, the  spatial infinity, $ar\rightarrow\infty$,
corresponds in fact to two degenerated spatial infinities
$x\rightarrow\infty$.
The Penrose diagram is very simple. The
causal structure of this manifold may then be interpreted in the usual
way, like the Schwarzschild solution. However the future infinity is
timelike instead of null.

The black hole solution (14) and (15) is asymptotic to the linear dilaton
spacetime (12) and (13). Now, asymptotically there is a timelike
killing vector $\partial\over{\partial t}$. Thus we expect the existence of the
conserved quantity $T_{0a}\xi^a$, where $\xi^a$ is the normalized killing
vector.
Using standard techniques one can calculate the ADM mass at
$x\rightarrow\infty$,
given by,
$$M_{total}={\lambda\over\sqrt{6}}e^{-3\phi_0}, \eqno(17)$$
where $\phi_0$ is the value of the dilaton at $x=0$, (see equation
(14)). We are considering here $\lambda^2>0$.
The ADM mass can also be calculated in the Schwarzschild gauge,
giving,
$$M_{total}={\lambda\over\sqrt{6}}b. \eqno(18)$$
Thus one can identify $e^{-3\phi_0}=b$.
Analogously to the 2D string theory, the mass of the black hole is
linked with the value of the dilaton at the horizon.

In order to include quantum field effects in the classical
geometry of the black hole we must compute the Hawking temperature. The
Euclideanized solution is periodic in imaginary time with period
${2\pi\over{\sqrt{3\over2}\lambda}}$. This is characteristic of a
thermal state emitting radiation at temperature,
$$T=\sqrt{3\over2}{\lambda\over{2\pi}}.\eqno(19)$$
This is independent of the mass, a result which has been reported in
other 2D theories, and which we will be interpreting below in the
context of a 4D planar spacetime. A step further towards a full
quantum gravitational treatment should include back reaction of the
radiation on geometry by using the trace anomaly relation [3].

How can we relate these solutions with the 4D world? By construction, solutions
(14) and (15) can be paste together to yield,
$$ds^2=-\tanh^2\left(\sqrt{3\over2}\lambda x\right) \cosh^{4\over3}\left(
\sqrt{3\over2}\lambda x\right)dt^2 + dx^2$$
$$\quad\quad\quad\quad +\cosh^{4\over3}\left(\sqrt{3\over2}\lambda x\right)
e^{-2\phi_0} \left( dy^2 + dz^2\right), \eqno(20)$$
which is a planar black hole in General Relativity!

A mass $M$ in 2D is a surface density $\sigma$ in 4D. From (6) we can
infer that they are connected through $M\sim \sigma e^{-2\phi}$. Now,
if we  increase $-\phi_0$, we are changing the gauge of  our planar
coordinates, $y\rightarrow e^{-\phi_0}y$, $z\rightarrow e^{-\phi_0}z$,
or in other words, our units of length are being decreased. An aereal
unit is transformed into a smaller area, so $\sigma$ increses, and therefore
$M$ increases. This 4D viewsight clears up the dependence of the 2D mass on
$\phi_0$. We would like to stress that solution (20) is not a domain wall,
it is indeed a black hole, as we further discuss below.

The independence of the Hawking temperature on the 2D mass can also be
explained.  Changing our units of length on the planar 2-surface
doesnot alter the units on the orthogonal spacetime surface.  Thus the
event horizon in Schwarzschild-like  coordinates is at
$r_H=\sqrt{3\over2}{1\over\lambda}$, and its surface gravity is
$k=\sqrt{3\over2}\lambda={3\over{2r_H}}$. Since these are not altered
by such a change of units, there is nothing to change the 2D
temperature. A crude dimensional argument shows that a 4D mass $m$,
must be linked to $\lambda$ by $m={1\over{\alpha\lambda}}$ where
$\alpha\geq0$ is a constant.  Therefore in the 4D world one would have
$T\sim {1\over m}$, restoring our expectations.

We have found and commented on two solutions only, namely the free
dilaton and the black hole. But, of course, there are many other
solutions. Equations (6) and (7) have an in-built symmetry which
transforms vacuum static solutions into homogeneous (time-dependent)
solutions by making, $\lambda\rightarrow i\lambda$, $t\rightarrow
i\chi$ and $x\rightarrow i\tau$. In addition, from 4D planar General
Relativity we know we have the Taub, the planar Kasner, and the
Horsk\'y-Novotn\'y solutions [10], which are also solutions in this 2D
theory. All of these have interesting causal structures.  Out of these,
the most fundamental is maybe Taub's which has so far eluded a clear
interpretation. In both backgrounds, either in 2D or in 4D planar
symmetry, we can now interpret the Taub solution as the spacetime which
takes over when the geometrical singularity of the planar black hole is
approached. In this context, one is carried into the viewpoint  that
the Taub solution is an approximation to the Schwarzschild spherical
solution near the singularity, inside the event horizon. Although the
topology of the 2D spatial surfaces of the orbits of the group of
isometries as well as the groups themselves are very different, (in
the planar case being $S^1{\rm x}S^1$ or $R^2$ with the group being
composed of two translations and one rotation, and in the spherical
case being $S^2$ with the $SO\left(3\right)$ group), an observer in that
region could set up some semi-local coordinate system, (not as local as a
quasi-Minkowskian observer, at least on a trajectory orthogonal to the
2D spatial surfaces), where Taub spacetime approximates the
Schwarzschild solution.

Another interesting solution in this 2D theory is the one-particle
(delta-function) solution with horizons [13], which corresponds in 4D
to the Ipser dust domain wall [14], and which differs from the black
hole given in (20). In this connection we mention that a
solution with horizons in a 2D theory was found by Brown,
Henneaux and Teitelboim [15].  It is a one-particle solution and
relates to the Vilenkin wall [16]. It is an object which doesnot belong
to the theory we have been presenting, since the theory itself cannot
admit `transversal' pressures. In fact this particle solution is an
object of the $R=T$ theory in 2D [17,18,8]. One could continue to list
several other possible 2D solutions containing matter: the multiple
particle solution [19], the string (cosmological) solution, and the
`beadcollar' string solution (i.e., a string sprinkled with particles)
[13]. One could also try gravitational collapse in 2D. A relation
between the collapse of dust and the collapse of null radiation could
be found, as it is suggested by the spherical collapse in 4D [20]. This
would also allow to test cosmic censorship and the formation of naked
singularities.

String theories in 1+1 dimensions are being used to gain insight
towards a quantum treatment of the graviton. One problem that is always
raised is, how the 2D results connect to the 4D world. In this letter
we have built a bridge that provides such a link. We have
constructed a 2D theory, with a structure similar to the string theory,
which is formally and directly related to 4D planar symmetry in General
Relativity. The other part  emerges  when we are able to relate generic
features, (such as the existence of black holes), from within these
type of 2D theories, using the Brans-Dicke parameter $\omega$ [21].

\bigskip\bigskip
\noindent Acknowledgements- I thank useful conversations with Jos\'e
Mour\~ao.  I have profited from the lectures on two-dimensional systems
given by V. Frolov  in the First Iberian Meeting on Gravity at \'Evora,
Portugal, in September 1992.  Research grants from
CNPq-Brazil and JNICT-Portugal are acknowledge as well as
support to travel from GTAE. I am also grateful to
Centro de F\'{\i}sica da Mat\'eria Condensada-IFM,
Lisbon, for providing the space and material
facilities  to initiated
and finished this work during the northern hemisphere Winter of 1992/93.
This work received a honorable mention of the Gravity Research Foundation
awards in 1993.
\vfill\eject
\centerline{References}
\medskip

\item{1.} G. Mandal, A. M. Sengupta, S. R. Wadia, {\it Mod. Phys. Lett. A},
{\bf 6}, 18, (1991).
\item{2.} E. Witten, {\it Phys. Rev. D}, {\bf 44}, 314, (1991).
\item{3.} C. G. Callan, S. B. Giddings, J. A. Harvey, A. Strominger,
{\it Phys. Rev.  D}, {\bf 45}, R1005, (1992).
\item{4.}  J. G. Russo, L. Susskind, L.
Thorlacius, {\it Phys. Lett. B}, {\bf 292}, 13, (1992).
\item{5.}  S. W. Hawking,
{\it Phys. Rev. Lett.}, {\bf 69}, 406, (1992).
\item{6.} C. Teitelboim, in {\it Quantum Theory of Gravity, essays in honor of
 the 60th birthday of B. DeWitt}, ed. S. Christensen, p. 327, Adam
Hilger-Bristol, (1984).
\item{7.} R. Jackiw, in {\it Quantum Theory of Gravity,
essays in honor of the 60th birthday of B. DeWitt}, ed. S. Christensen,
p. 403 , Adam Hilger-Bristol, (1984).
\item{8.} J. P. S. Lemos, Paulo M. S\'a, ``The Two-Dimensional Analogue
of General Relativity'', {\it Class. Quantum Gravity}, in press, (1993).
\item{9.} T. Banks, M. O'Loughlin, {\it Nucl. Phys. B}, {\bf 362}, 649,
    (1991).
\item{10.} D. Kramer, H. Stephani, M. MacCallum, E. Herlt, {\it Exact Solutions
of Einstein's Field Equations}, Cambridge University Press, (1980).
\item{11.} Yu. A. Kubyshin, J. M. Mour\~ao, G. Rudolph, I. P. Volobujev,
{\it Dimensional Reduction of Gauge Theories, Spontaneous Compactification and
Model Building}, Lectures Notes in Physics vol. 349, Springer-Verlag, (1989).
\item{12.} P. Thomi, B. Isaak, P. Hajicek, {\it Phys. Rev. D}, {\bf 30},
1168, (1984).
\item{13.} J. P. S. Lemos, (to be published).
\item{14.} J. R. Ipser, {\it Phys. Rev. D}, {\bf 30}, 2452, (1984).
\item{15.} J. D. Brown, M. Henneaux, C. Teitelboim, {\it Phys. Rev. D},
{\bf 33}, 319, (1986).
\item{16.} A. Vilenkin, {\it Phys. Lett. B}, {\bf 133}, 177, (1983).
\item{17.} R. B. Mann, A. Shiekh, L. Tarasov, {\it Nucl. Phys. B} {\bf 341},
134, (1990).
\item{18.} R. B. Mann, S. F. Ross, {\it Class. Quantum Grav.}, {\bf 9}, 2335,
   (1992).
\item{19.} P. S. Letelier, {\it Class. Quantum Grav.} {\bf 7}, L203, (1990).
\item{20.} J. P. S. Lemos, {\it Phys. Rev. Lett.}, {\bf 68}, 1447, (1992).
\item{21.} J. P. S. Lemos, P. M. S\'a, {\it Phys. Rev. D}, {\bf 49}, 2897
(1993).

\end